# Auditing Asset-Specific Preferences in Financial Large Language Models

*Evidence from Bitcoin Representations and Portfolio Allocation*

Wenbin Wu

Cambridge Centre for Alternative Finance, University of Cambridge

May 2026

## Abstract

Large language models now power robo-advisors and trading agents, yet whether they carry built-in biases toward specific assets is largely untested. We ask three questions: do LLMs systematically prefer certain financial instruments; can an internal representation with causal leverage over those preferences be identified; and does that representation affect downstream financial decisions?

We develop a three-level audit protocol and apply it to Bitcoin. First, a behavioral audit of nine frontier LLMs shows that Bitcoin's ranking among money-like instruments is frame-dependent: models place it around rank 5 of 8 as "reliable money" but near the top under crisis and autonomous-agent frames, and an attribute-swap experiment shows that rankings track functional properties, not names. Second, we open a model's internals: a search across thousands of sparse-autoencoder features in Gemma 3 identifies a dominant Bitcoin-selective feature. Amplifying it shifts the model toward the asset and suppressing it shifts the model away, even when "Bitcoin" never appears in the prompt. Third, we test financial consequences: amplification raises Bitcoin's portfolio share by 5.2 percentage points while suppression lowers it by 4.6 pp, with amplification reallocating within crypto and suppression cutting total crypto exposure.

We characterize this as *bounded behavioral leverage* (leverage meaning causal influence over outputs, not financial leverage): an identifiable internal feature can be perturbed to move financial choices, but only within measurable limits. The framework links internal representations to external recommendations, validated with random controls and mechanism boundaries. As LLMs become autonomous financial agents, this is a first step toward a behavioral layer for emerging know-your-agent (KYA) standards: knowing what an agent prefers, and how far that preference can be moved.

# 1 Introduction

Large language models now power robo-advisory platforms, financial search engines, and autonomous trading agents that collectively influence portfolio decisions at scale (Dong et al., 2025; Li et al., 2023); equity trading volume drops during ChatGPT service outages, direct evidence that investors already route real trading decisions through these tools (Cheng et al., 2025). When these systems compare instruments or frame investment trade-offs, any systematic asset-level bias becomes a model-risk and consumer-protection concern.

Regulators already require that algorithmic advice be fair and explainable, and international bodies are now extending oversight to AI specifically: IOSCO's 2026 supervisory toolkit covers AI in capital markets across the full lifecycle, including emerging agentic AI (International Organization of Securities Commissions, 2026), while central banks warn that poorly-controlled AI amplifies financial-stability risks (Zhang, 2026).

A fast-emerging *know your agent* (KYA) agenda, the agentic analog of know-your-customer due diligence, focuses on verifying an agent's identity, authority, and permissions (Chaffer, 2025; Grogan, 2025). But identity-layer due diligence does not say how an agent behaves with money: whether it favors particular assets, how far that preference bends with framing, and whether it can be deliberately moved. Turning these principles into concrete supervision demands audit tools that go beyond surface-level prompt testing, consistent with work on technology-enabled supervision and accountability in financial AI (Arner et al., 2017; Buckley et al., 2021; Zhang et al., 2019).

Cryptoassets sharpen the problem because they span financial innovation, payment infrastructure, and regulatory oversight (Rauchs et al., 2018). A recent Bitcoin Policy Institute (BPI) report (Danielian et al., 2026) claimed that AI agents "overwhelmingly prefer Bitcoin and digital-native money over traditional fiat," testing 36 models under an autonomous-agent prompt. A financial regulator or model-risk team cannot take that finding at face value without knowing whether it reflects a prompt frame, a training-data artifact, or an internal representation that could be detected and bounded. Answering that question requires an audit protocol connecting surface outputs to internal structure to downstream financial decisions.

This paper takes a step toward *behavioral KYA*: due diligence on what an agent actually prefers. Throughout, *preference* denotes an operational, revealed measure, the model's relative choice tendency under a given frame and choice set, not a stable utility function in the classical sense. It builds that protocol and applies it to monetary-asset preferences, using Bitcoin as a case study. The paper makes three contributions summarised in Figure 1.

1. **Financial AI audit protocol.** A framework linking output preference, internal representation, and downstream decision effects. The protocol produces a *representation–leverage assessment*: whether a discovered internal representation has validated leverage over constrained financial choices.

2. **Behavioral audit.** Cross-model ranking of 8 money-like instruments across 8 financial frames, with attribute-swap experiments separating name recognition from functional profiles. Rankings prove sharply frame-dependent: Bitcoin ranks near the bottom under "reliable money" but near the top under crisis and autonomous-agent frames.

3. **Bounded behavioral leverage.** A differential activation search across thousands of sparse-autoencoder (SAE) features at four Gemma 3 scales identifies Bitcoin-selective repre-sentations at every scale, with steering leverage strongest at 27B. Downstream allocation tests show that positive decoder-direction injection raises Bitcoin allocation while negative injection lowers it, with both effects bounded and frame-contingent.

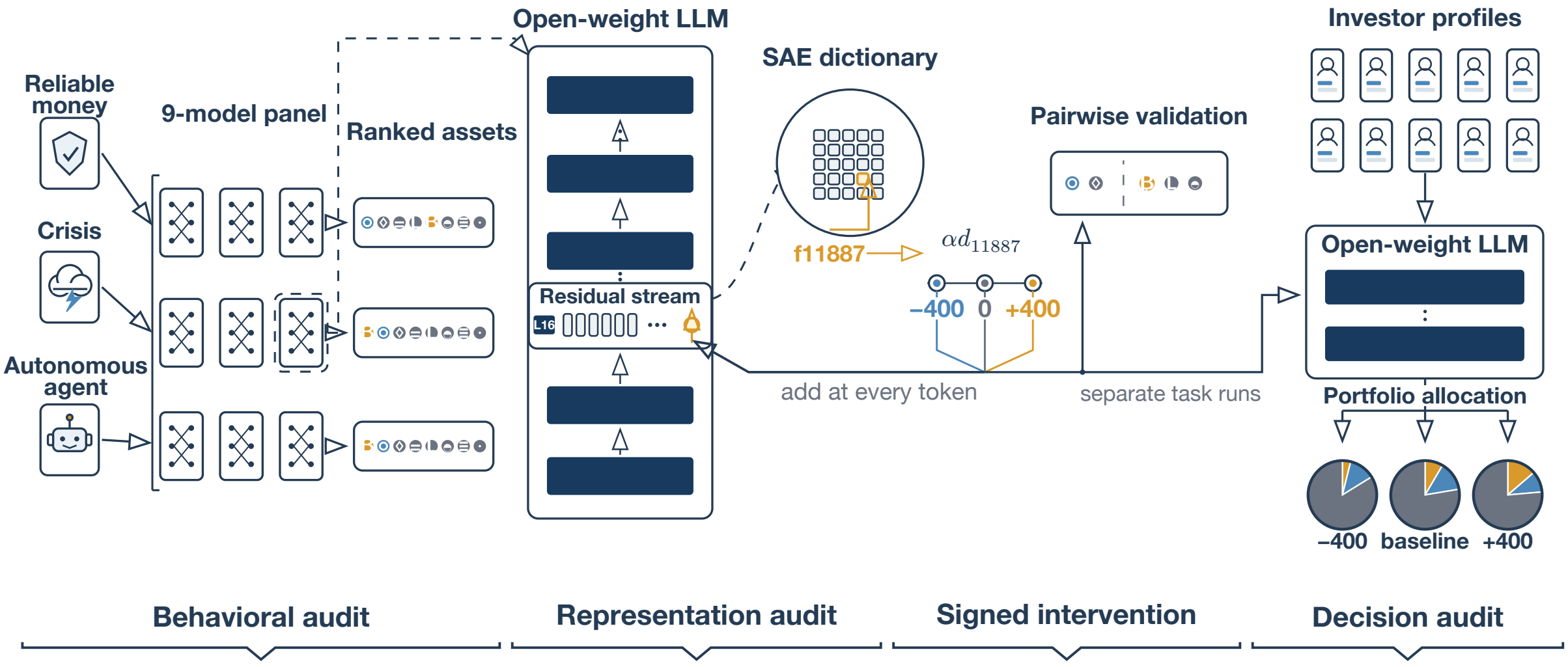

Figure 1: Audit design and evidence flow. A nine-model panel measures frame-dependent asset rankings (left). The internal audit then localizes a Bitcoin-selective SAE direction in Gemma 3 27B and tests signed decoder-direction injections on pairwise choices (center). The same intervention is evaluated in profile-conditioned portfolio allocation (right), where sectors show Bitcoin, other crypto, and other assets using the exact aggregate means. Dashed lines denote diagnostic selection or inspection rather than causal flow; solid intervention arrows begin at the signed injection.

**Related literature and positioning.** This paper bridges three literatures that remain disconnected: LLM preferences in financial settings, activation steering and SAE interpretability, and financial AI governance.

Large language models can be treated as economic agents whose preferences are elicitable and measurable (Chen et al., 2023; Filippas et al., 2024). Ouyang et al. (2024) find stable yet diverse risk profiles across 50 LLMs; Liu et al. (2025) and Song et al. (2025) show that measured preferences shift with persona and prompt language. Lee et al. (2025) document model-specific investment biases; Fieberg et al. (2025) audit portfolio advice across 32 LLMs; Wang & Gu (2026) develop fairness frameworks for demographic bias in LLM advice; and expert-level cryptocurrency benchmarks document large cross-model capability differences on crypto-analysis tasks (Guo et al., 2025). No existing work tests whether these output-level preferences reflect identifiable internal representations or are emergent artifacts of distributed computation.

On the interpretability side, Turner et al. (2024) introduce contrastive activation addition; Templeton et al. (2024) demonstrate that individual SAE features can govern concept-specific outputs (the "Golden Gate Bridge" feature in Claude 3 Sonnet); Mayne et al. (2024) decompose steering vectors into SAE features; and Arad et al. (2025) show that activation selectivity does not imply steering leverage. Additional methods include targeted linear approximation (Chalnev et al., 2024), multi-layer Gaussian schedules (Góral et al., 2025), circuit-level pruning (Gao et al., 2025), and steering vectors recovered by aligning behaviorally elicited preferences with neural representations, applied to risk preferences (Zhu et al., 2025).

In financial AI governance, Chen et al. (2025) introduce a "Financial Brain Scan" using SAE features to map broad financial reasoning concepts. The broader governance context includes

AI adoption in financial services (Dong et al., 2025; Li et al., 2023), robo-advice (Piehlmaier, 2022), emerging audit frameworks (Hu et al., 2023; Kang et al., 2025; Kou & Lu, 2025), a fast-growing literature on agentic AI in finance and its governance (Gong, 2026; Kurshan et al., 2025; Meng & Chen, 2026), and evidence that LLM outputs vary substantially with framing (Brucks & Toubia, 2025; Santurkar et al., 2023). The monetary-theory foundations motivate both the attribute profiles and the ranking task itself. Classical accounts establish the properties of money (Jevons, 1875) and the evolutionary, liquidity-based origin of moneyness (Menger, 1892), while the money-view tradition treats money as an inherent hierarchy of instruments graded by moneyness (Mehrling, 2013; Young, 1999), the ordering our cross-model "reliability as money" ranking measures directly. Recent work extends this hierarchy to cryptoassets and decentralized finance (Wu, 2026).

The remainder of the paper proceeds as follows. Section 2 presents the behavioral audit: cross-model rankings, frame sweep, and attribute-profile decomposition. Section 3 opens the model internals, identifying and validating a Bitcoin-selective SAE feature, then bounding what that feature is mechanistically. Section 4 tests downstream financial decision effects. Section 5 gathers limitations and implications for financial AI governance.

# 2 Behavioral Audit of Monetary-Asset Judgments

## 2.1 Design

The behavioral audit compares eight instruments, namely US dollar cash, bank deposit, US Treasury bill, gold, Bitcoin, Ethereum, S&P 500 index fund, and residential real estate, across nine contemporary LLMs: Claude Opus 4.7, GPT-4.1, Gemini 2.5 Pro, Grok 4.3, Gemma 4 31B, Gemma 3 27B IT, DeepSeek V4 Pro, Qwen3 235B, and Llama 4 Maverick. Each instrument carries a fixed binary profile across 8 attributes (full design matrix in Table 13). Bank Deposit and Treasury share identical profiles, so any rank difference between them must arise from label identity or uncoded attributes.

Five experimental conditions form a 2×2 factorial plus one synthetic-label control (Table 1). The factorial crosses label type (real names vs. anonymous "Instrument 1–8") with attribute assignment (real profiles vs. BTC↔Treasury profiles swapped). A fifth condition replaces "Bitcoin" with the novel label "Bitstone" while keeping Bitcoin's real attribute profile.

| | **Real profiles** | **Swapped profiles** |
|---|---|---|
| **Real names** | A: baseline | C: profile effect |
| **Anonymous** | B: label effect | D: both removed |
| **E**: Synthetic label "Bitstone" + Bitcoin's real profile (name-independence check) | | |

Table 1: Experimental conditions. A vs. B isolates the label contribution; A vs. C isolates the profile contribution; E tests whether a novel label with Bitcoin's attributes ranks like Bitcoin.

Each model ranks the 8 instruments "from most to least reliable as money" in 10 repetitions per condition, with randomized instrument order and temperature = 0. Two independent runs yield 833 parsed trials from 900 evaluations. To test frame sensitivity, the same design was rerun across 8 financial frames: *reliable money*, *store of value*, *medium of exchange*, *investment return*, *portfolio diversifier*, *crisis/capital controls*, *reserve asset*, and *autonomous economic agent*, with 10 deterministic repetitions per model-frame cell. Parse-success rates are reported in A.3.

## 2.2 Stable aggregate ordering

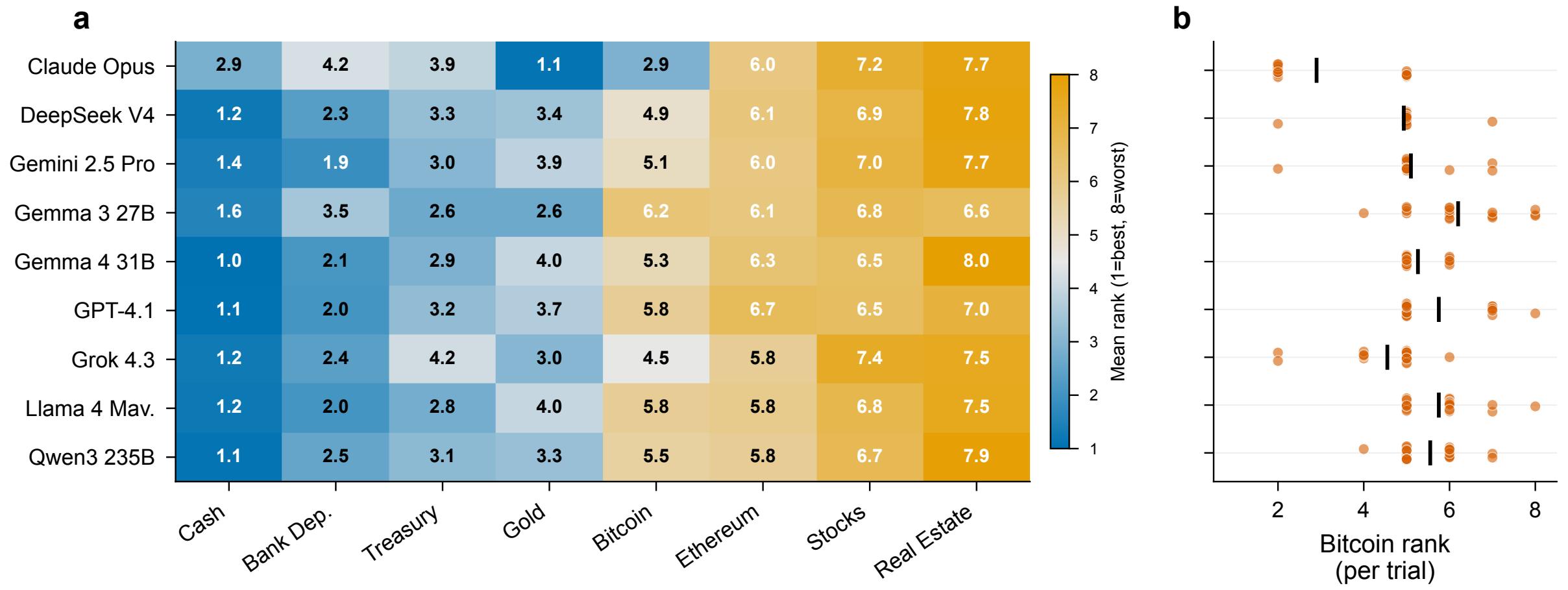

Figure 2: **(a)** Cross-model ranking under the named_real condition ($n = 174$ parsed trials, pooled over two independent runs per model). Each cell shows one instrument's mean rank for that model; lower rank = more reliable as money. **(b)** Per-trial Bitcoin rank for each model (temperature = 0; each dot is one trial, black bar is the mean).

Figure 2 **(a)** displays the aggregate ranking. State-backed instruments, namely Cash, Bank Deposit, and Treasury, occupy the top positions, followed by Gold, then the digital and real assets. Cross-model concordance is high and statistically significant, measured by Kendall's coefficient of concordance ($W = 0.92$ over the per-model orderings of Figure 2, $p < 0.001$), and tightens to $W = 0.96$ once the Claude Opus outlier is excluded.

Figure 2 **(b)** shows per-trial Bitcoin ranks. Claude Opus 4.7 is the only model whose mean Bitcoin rank falls in the top 3; the remaining eight cluster in the bottom half. The narrow per-trial spread confirms that temperature = 0 produces near-deterministic outputs, with variation driven by prompt-order sensitivity only.

The frame sweep (Section 2.4) reveals that Claude's elevated Bitcoin ranking concentrates in investment-oriented frames; under *reliable money* and *reserve asset*, Claude is unremarkable. An extended cross-provider panel (A.1) shows that Claude Opus 4.7 and Claude Sonnet 4.6 behave alike, both ranking Bitcoin near the bottom under a money frame and at the top under a long-horizon frame, so the pattern is family-level rather than a single-model quirk. Several non-Claude models elevate Bitcoin under the long-horizon frame too, so the effect is not Claude-specific, and a shared training or RLHF lineage is one plausible driver (Santurkar et al., 2023). Because both Claude models are closed-weight, testing this at the representation level requires open-weight access, precisely the diagnostic gap that motivates the representational audit that follows.

## 2.3 Replication and format robustness

The aggregate ranking replicates across independent runs and response formats. An independent replication preserves Bitcoin's aggregate position (rank 5 of 8 in both runs) and reproduces the per-model orderings exactly for four of the nine models; the remaining differences are reorderings among closely ranked instruments, with per-instrument mean-rank shifts of at most 1.1. Pairwise forced-choice framing on Gemma 3 27B confirms that baseline prefers every traditional asset over Bitcoin, providing format-robustness evidence beyond list-ranking effects. Full per-model rankings appear in A.2.

## 2.4 Within-study frame sweep

The reliable-money ranking answers one financial question. To test how sensitive the result is to framing, the frame sweep tests the same nine models and eight instruments under eight financial frames. Table 2 reports Bitcoin's mean rank and top-3 rate by frame.

| Frame | Parsed | BTC rank | Top-3 | BTC win vs Treasury |
|---|---|---|---|---|
| Reliable money | 89 | 5.01 | 9.0% | 10.1% |
| Store of value | 89 | 5.38 | 24.7% | 19.1% |
| Medium of exchange | 90 | 3.57 | 57.8% | 90.0% |
| Investment return | 88 | 3.02 | 77.3% | 88.6% |
| Portfolio diversifier | 87 | 3.10 | 65.5% | 69.0% |
| Crisis / capital controls | 87 | 1.51 | 96.6% | 96.6% |
| Reserve asset | 90 | 6.11 | 2.2% | 1.1% |
| Autonomous economic agent | 89 | 1.20 | 100.0% | 100.0% |

Table 2: Deterministic nine-model frame sweep. Same model panel, same 8 instruments, same ranking response format, 10 repetitions per model-frame cell. Lower rank is more favorable.

As Table 2 details, under *reliable money* and *reserve asset*, Bitcoin ranks in the bottom half, rarely beating traditional instruments. Under *crisis/capital controls* and *autonomous economic agent*, it rises to near the top. The top-3 rate spans nearly the full range across frames. This frame dependence reconciles the reliable-money result with BPI-style agentic findings (Danielian et al., 2026) without treating either as the model's unconditional preference. Grok 4.3, which leads expert-level cryptocurrency-task benchmarks (Guo et al., 2025), shows the sharpest version of the pattern, ranking Bitcoin 4.6 under reliable money but 1.0 under both the crisis and autonomous-agent frames: capability on cryptocurrency tasks does not imply preference for cryptoassets as money.

## 2.5 Label vs. attribute-profile decomposition

The five conditions in Table 1 separate how much of Bitcoin's ranking comes from its *name* versus its *attribute profile*. Because attributes are bundled by instrument (rank-deficient design, rank 7/8), the comparisons capture profile-level associations, not isolated attribute effects (rank diagnostics in Appendix B). At temperature = 0, within-model variation arises solely from randomized instrument order; the bootstrap intervals below quantify *order-sensitivity*, not sampling uncertainty.

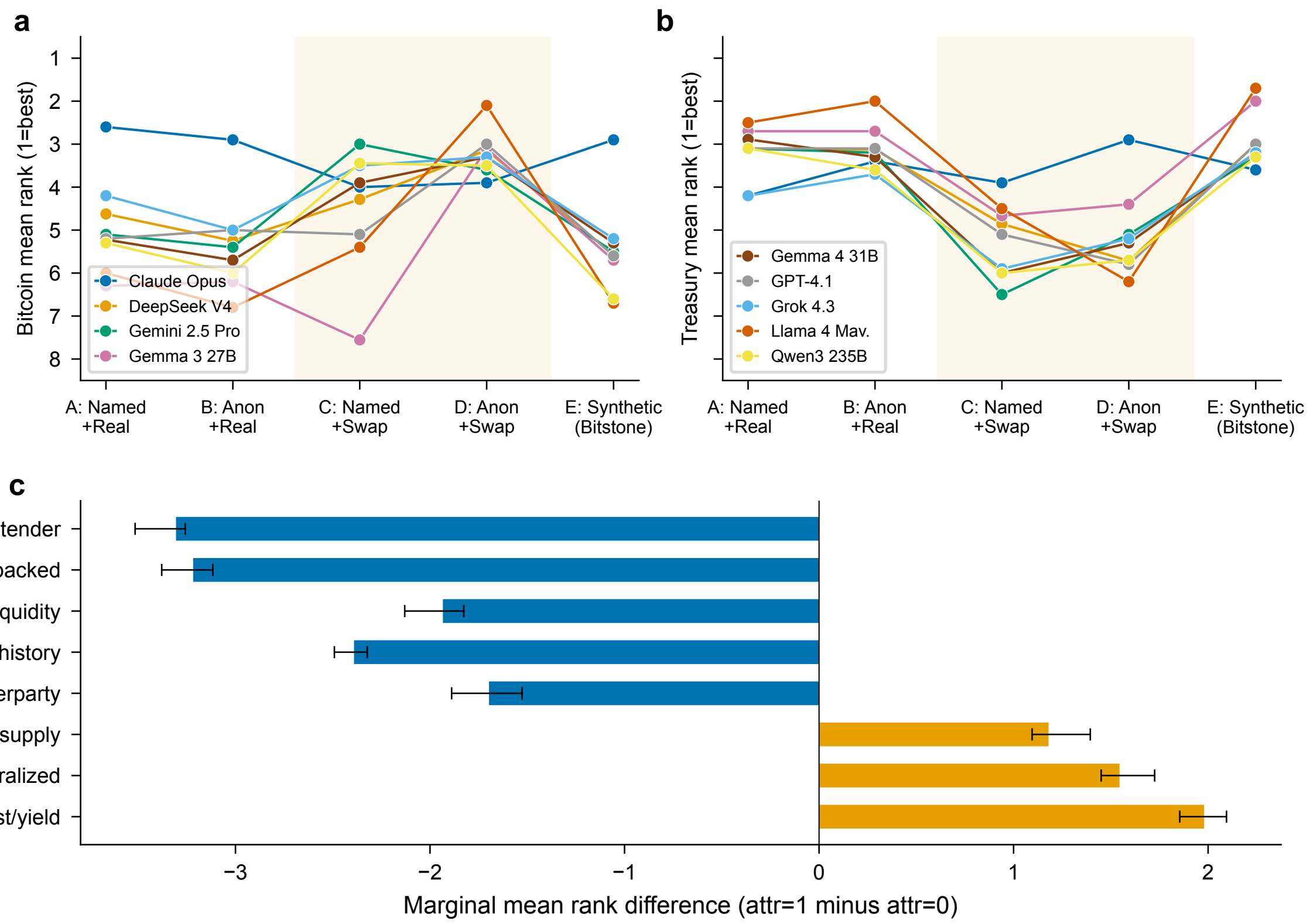

Figure 3: Profile-swap experiment and attribute decomposition ($n = 434$ parsed trials, v2 confirmatory run, temperature = 0). **(a, b)** BTC and Treasury mean rank across 5 experimental conditions for each of 9 models; orange shading marks swap conditions; the model legend is split across panels (a) and (b); condition E uses the synthetic label "Bitstone" with Bitcoin's attribute profile. **(c)** Marginal attribute-profile associations with monetary-reliability rank. Negative values indicate more favorable (lower) rank. Blue marks attributes associated with more favorable ranks; orange marks attributes associated with less favorable ranks. Error bars show bootstrap 95% CIs over the named-real, anonymous-real, and synthetic-label conditions ($n = 262$).

Figure 3(a, b) displays the full condition matrix. The label effect is small, inconsistent in sign across models, and never exceeds one rank position; its pooled bootstrap interval contains zero. The profile-swap effect is several times larger: when Bitcoin receives Treasury's attribute profile under condition C, its rank improves by up to two positions depending on the model, while Treasury moves downward under Bitcoin's profile.

### 2.5.1 Attribute-profile associations

Figure 3(c) displays the marginal profile-bundle associations. Attributes associated with state-backed instruments, such as legal tender, government-backed, high liquidity, and long history, receive more favorable ranks, while attributes associated with cryptoassets, such as decentralized and fixed/scarce supply, receive less favorable ranks. All nine models agree on the sign direction for legal tender, government backing, and long history (eight of nine for high liquidity), and eight of nine agree on the crypto-associated attributes. The attribute design matrix is rank-deficient, so these are profile-bundle associations, not isolated attribute effects; rank diagnostics are in Appendix B.

### 2.5.2 The "Bitstone" test

When "Bitstone" (a synthetic label unlikely to appear as a standard monetary instrument in training data) carries Bitcoin's attribute profile, it ranks similarly to Bitcoin (condition E vs. A), confirming that the attribute-profile bundle drives rank more than the instrument label alone.

Taken together, the profile-swap, attribute-decomposition, and Bitstone results indicate that Bitcoin's low ranking is attribute-driven. Models penalize the combination of decentralization and fixed supply, and reward the combination of legal-tender status, government backing, and high liquidity. The instrument name contributes little.

## 3 Internal Representations and Their Causal Leverage

Behavioral rankings tell us *what* models output but not *why*. If a model consistently ranks Bitcoin low, the cause could be a specific internal representation or an emergent property of distributed computation with no identifiable locus. Answering this question matters for governance: a localized, auditable representation implies different regulatory tools than a diffuse, uninterpretable one.

This section opens the model and asks whether Bitcoin-related outputs have detectable internal correlates and whether nudging those correlates shifts the model's asset-choice logits (the raw output scores it produces before they become probabilities). The approach is analogous to factor analysis in finance: just as a portfolio's returns can be decomposed into loadings on interpretable factors, a neural network's internal activations can be decomposed into loadings on sparse dictionary directions, some of which correspond to identifiable concepts. Sparse autoencoders (SAEs) provide this decomposition: an SAE is a small network, trained on the model's internal activity, that rewrites that activity as a short list of separately interpretable features, each one a single direction. When a direction is identified, "steering" injects or suppresses that direction during the model's computation, analogous to asking: what happens to portfolio behavior if we artificially increase or decrease the loading on a single factor?

### 3.1 Model, SAE setup, and methodological progression

The analysis uses GemmaScope 2 SAEs (McDougall et al., 2025), which provide 16,384 JumpReLU features per layer (Lieberum et al., 2024; Rajamanoharan et al., 2024), applied to Gemma 3 27B IT at layers 16, 31, 40, and 53. Full feature identifiers, hook points, and SAE widths for cross-model replication are in Table 15.[1] These SAEs are trained with a Matryoshka-style objective designed to prevent feature absorption (Chanin et al., 2024), so the absorption failure mode that complicates feature-level analysis in earlier SAE generations does not arise here.

Two design choices follow from how the layer-16 SAE behaves under perturbation. First, injecting 200 random SAE directions at high scale shows that layer 16 is broadly sensitive: over 80% of random directions shift BTC rank, compared with under 10% at other layers. Rank movement alone is therefore not evidence of content specificity; the operative question is whether a direction produces effects that random directions do not, which is what the random-control baselines below test.

Second, the feature is located bottom-up, by differential activation. The intuitive interpretability route, taking a working steering vector and decomposing it into SAE features to read off the concept it exploits, does not isolate a Bitcoin feature here: the Mayne et al. (2024) decomposition of a composite steering vector loads almost entirely on *dense* features that fire on essentially all text regardless of category (generic components of the residual stream, the running vector the transformer passes from layer to layer, rather than monetary content), while the Bitcoin-selective feature carries near-zero weight. This is the input-feature/output-feature distinction of Arad et al. (2025): steering vectors are built from steering-effective *output* features, whereas the concept we want to audit is an activation-selective *input* feature, and the two rarely coincide. We therefore search the full dictionary bottom-up for the feature that responds selectively to Bitcoin content, then test its leverage separately.

[1]GemmaScope 2 denotes sparse autoencoders trained on the Gemma 3 family; the version index refers to the SAE release, not the model generation.

A training-free baseline makes the same point from the demonstration side. Following the in-context-vector recipe (Liu et al., 2024) in a single-layer instantiation matched to our injection point, we average layer-16 last-token activation differences over twelve contrastive advice pairs whose completions recommend or reject Bitcoin with asset names, risk vocabulary, and length controlled. The resulting direction is nearly orthogonal to f11887 (cosine 0.005), and at norm-matched scales it is also behaviorally inert: injection leaves the pairwise generation battery at its baseline (11 of 14 Bitcoin wins, unchanged at both scales; under this strict-parse generation protocol the baseline favors Bitcoin more often than in the log-probability battery), and the reversed direction shifts it only to 9 of 14, where f11887 suppression leaves 4 of 14 single-order generations and no pair won in both orders. Geometric misalignment alone would not settle much, since the decomposition above already shows that low cosine can coexist with shared steering effects; the informative part is that the contrastive-context direction fails on both axes at once, neither aligning with the audit object nor reproducing its leverage. Demonstration averaging is a probe of behavior, not a substitute for locating the representation that carries it.

### 3.2 Differential activation search: Identifying Feature 11887

The search for Bitcoin-selective activation spans all 16,384 features in the Layer 16 SAE. Sixty prompts across 4 categories (15 each: Bitcoin-positive, crypto-general, traditional finance, non-financial) are passed through the model; L16 residual-stream activations are encoded through the SAE and features ranked by Cohen's $d$ between the Bitcoin-positive category and all others combined.

| Feature | Cohen's $d$ | BTC mean | Crypto | Trad. fin. | Non-fin. | BTC nz rate |
|---|---|---|---|---|---|---|
| **f11887** | **2.055** | **28.52** | 0.00 | 0.00 | 0.00 | **0.733** |
| f595 | 0.544 | 1.23 | 0.00 | 0.00 | 0.00 | 0.067 |
| f9138 | 0.544 | 0.89 | 0.00 | 0.00 | 0.00 | 0.067 |
| f642 | 0.394 | 3.46 | 2.50 | 1.27 | 0.00 | 0.333 |
| f153 | 0.369 | 50.86 | 56.21 | 33.20 | 0.28 | 0.800 |

Table 3: Top 5 features by Cohen's $d$ (Bitcoin-positive vs. all other categories, 60-prompt discovery set). Columns report $d$, mean activation on Bitcoin-positive prompts, mean activation on crypto-general prompts, mean activation on traditional-finance prompts, mean activation on non-financial prompts, and nonzero rate on Bitcoin-positive prompts.

As Table 3 shows, Feature 11887′s Cohen's $d$ exceeds the next-best feature nearly fourfold (3.8×), with zero mean activation on all non-Bitcoin categories. The search yields a clearly dominant Bitcoin-selective SAE feature within this dictionary, establishing the precondition for the causal intervention tests that follow.

To avoid overfitting, no prompt set is reused across discovery, validation, and evaluation; the discovery set (60 prompts) is used only to select f11887, and all subsequent tests use non-overlapping prompts. The binary BTC-vs-non-BTC calibration set yields a higher Cohen's $d$ ($d = 5.6$ for f11887) than the four-category discovery set; the two estimates are reported separately throughout.

### 3.3 Causal validation: steering moves monetary choice

f11887 produces clean bidirectional control over monetary-choice logits. Steering injects the SAE decoder direction at layer 16 during the forward pass. On order-balanced pairwise comparisons, amplification at scale 400 flips every baseline-losing comparison to a Bitcoin win; suppression drives Bitcoin to zero wins. The headline scale is commensurate with the feature's natural operating range: on held-out Bitcoin prompts, f11887's per-prompt peak activation averages 380.8 (range 278–433; Table 16). Because first-token logits can be confounded by tokenizer length differences, the headline test uses full-string log-probability with per-token normalization across 7 assets. Table 4 reports the results.

| **Condition** | **Wins** | **Mean margin** | **Δ** | **Per-token Δ** |
|---|---|---|---|---|
| Baseline | 2/7 | −2.31 | — | — |
| f11887 s250 | 4/7 | +1.20 | +3.51 | +2.89 |
| **f11887 s400** | **7/7** | **+2.99** | **+5.30** | **+4.22** |
| f11887 s600 | 7/7 | +5.13 | +7.44 | +5.53 |
| f11887 s770 | 7/7 | +6.14 | +8.45 | +6.10 |
| Suppress s400 | 0/7 | −5.73 | −3.42 | −3.26 |
| Suppress s770 | 0/7 | −5.87 | −3.56 | −3.64 |
| Combo s770 | 7/7 | +2.14 | +4.45 | +3.36 |

Table 4: Full-string log-probability margins (7 assets, order-balanced, per-token normalized). "Wins" = comparisons where the order-balanced Bitcoin margin is positive. Conditions span amplification and suppression at multiple scales, plus the composite vector at s770.

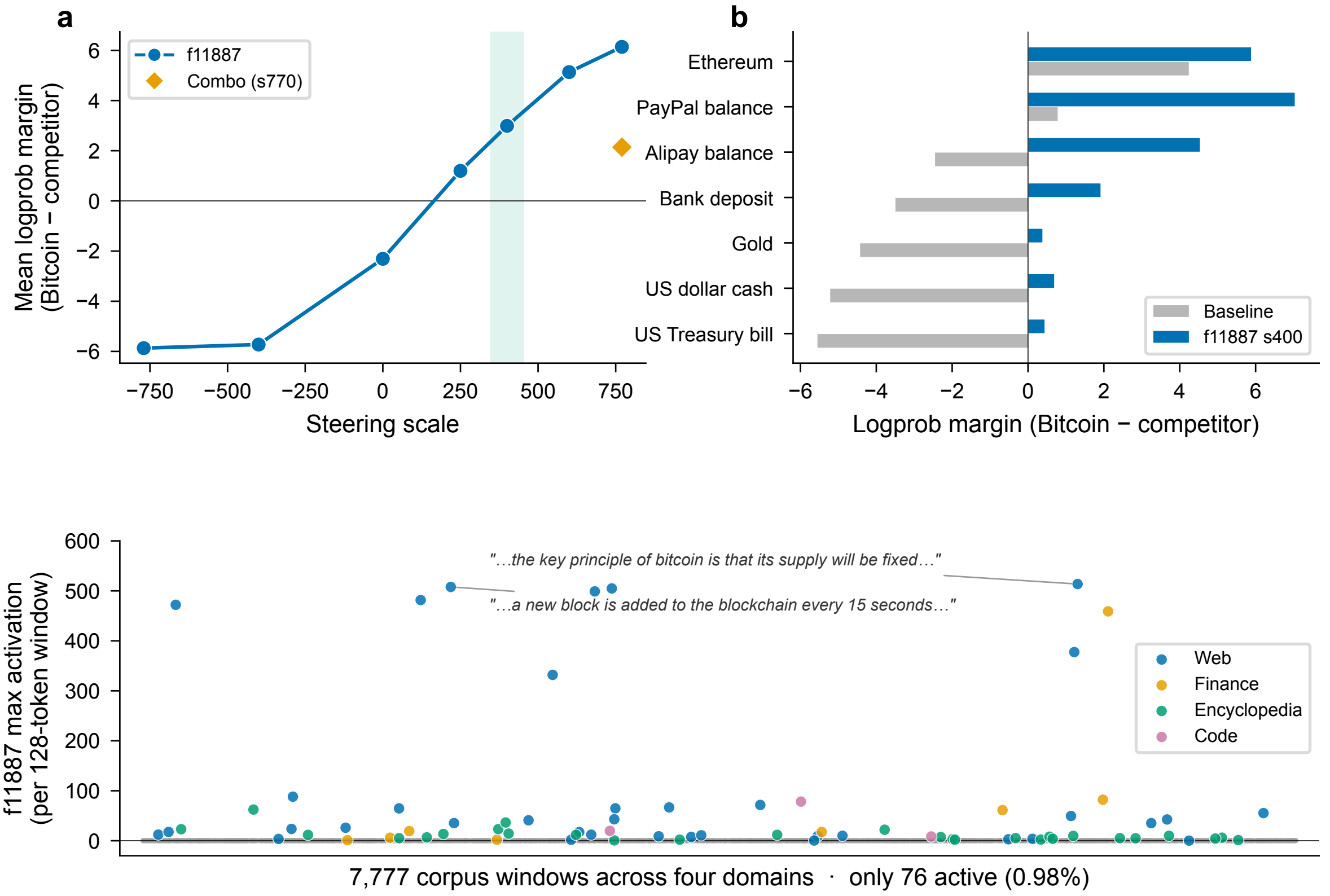

Figure 4: Bidirectional steering and corpus selectivity of f11887. **(a)** Dose–response: green band marks the scale range where pairwise wins and non-financial specificity are both maximal; combo vector at scale 770 shown for comparison. **(b)** Per-asset margins under baseline and amplification at scale 400; negative values indicate the model prefers the competitor. **(c)** Per-window maximum activation of f11887 across 7,777 windows (912{,} 624 tokens) sampled from four corpora; the feature is silent on all but 76 windows (0.98%), and every high-activation window is Bitcoin or blockchain text. See Table 7 for perturbation-as-percentage-of-residual-norm.

Figure 4(b) breaks down the steering effect by competitor asset. Under baseline, Bitcoin loses to every traditional asset (negative margins); at scale 400, amplification flips all seven comparisons to Bitcoin wins, with the largest gains against payment instruments (PayPal, Alipay) and the smallest against gold and Treasury. The effect extends to full rankings: on the 8-asset ranking task, f11887 moves Bitcoin monotonically from rank 6 at baseline to rank 1 at high scale, with format breakdown only at the highest scale tested.

## 3.4 Specificity and controls

Three families of test confirm that the steering effect is specific to Bitcoin content rather than a generic perturbation artifact. First, five non-financial ranking tasks (planets by size, countries by population, metals by density, animals by speed, languages by native speakers) test whether the feature disrupts general reasoning, with a validity criterion of $\geq 6$ output lines. Table 5 reports the specificity–effectiveness trade-off.

| Scale | Valid tasks | Pairwise wins | Assessment |
|---|---|---|---|
| s250 | 5/5 | 4/7 | Specific but incomplete |
| **s400** | **5/5** | **7/7** | **Sweet spot** |
| s600 | 3/5 | 7/7 | Pairwise intact, specificity degraded |
| s770 | 2/5 | 7/7 | Format collapse on non-financial tasks |

Table 5: Specificity–effectiveness trade-off. "Pairwise wins" = Bitcoin-vs-asset comparisons where the order-balanced Bitcoin margin is positive. "Non-financial valid" = non-financial ranking tasks that produce format-valid output. One task (metals by density) produces verbose but format-preserving output at s250 and s400.

Second, two selectivity tests confirm that f11887 is a Bitcoin-concept feature, not crypto-general or broadly financial. Across a 0.9M-token diverse corpus covering web, code, financial news, and encyclopedia text, f11887 is extremely sparse: only 76 of 7,777 windows are active (0.98%), and every high-activation window is Bitcoin or blockchain text, as Figure 4(c) shows. A held-out set of 95 prompts across 19 semantic categories confirms f11887 fires on all Bitcoin-related categories regardless of valence but does not fire on stablecoins, generic crypto, DeFi, fiat-critical content, Treasuries, or non-financial controls. The few non-Bitcoin activations are weak and isolated. Full per-category results are in Table 16.

Third, f11887′s steering effect survives a suite of 320 random controls spanning four families: random SAE features, decoder-norm-matched features, random Gaussian directions, and permuted-label nulls, all evaluated at scale 400 using the identical pairwise protocol. f11887 ranks at the 100th percentile across all families. An advice-format generation check ($n = 30$) gives a preliminary indication that the effect extends to unconstrained text generation, though Bitcoin is never the top recommendation; see Table 17. The 100th-percentile ranking across 320 diverse controls rules out generic-perturbation artifacts.

### 3.5 The anonymous instrument test: concept, not token

The preceding pairwise tests ask the model to compare "Bitcoin" against named assets. To test whether f11887 merely raises the probability of the string "Bitcoin," we presented the model with anonymous instruments described only by attribute profiles, with no asset names in the prompt or answer options.

Four profile pairs (Bitcoin-like vs. Treasury/Cash/Gold/Bank attributes) are tested across two to three prompt lengths, four neutral label systems, both label-to-profile assignments, and six steering scales from $-400$ to $+400$. The Bitcoin-like profile includes fixed supply, decentralized settlement, no central issuer, global transferability, high volatility, and no government backing. The model chooses between neutral labels; the word "Bitcoin" never appears. As Table 6 reports, f11887 amplification at scale 400 shifts all 9 anonymous conditions toward the Bitcoin-profile instrument, with a single reversal in 36 condition-level comparisons.

| **Profile pair** | **Prompt** | **Baseline** | **s400** | **Δ** |
|---|---|---|---|---|
| BTC vs Treasury | short | −7.42 | −4.55 | **+2.88** |
| BTC vs Treasury | long | −7.17 | −5.70 | +1.47 |
| BTC vs Treasury | reviewer | −7.08 | −6.28 | +0.80 |
| BTC vs Cash | short | −7.92 | −6.43 | +1.49 |
| BTC vs Cash | long | −7.70 | −6.46 | +1.25 |
| BTC vs Gold | short | −8.70 | −8.23 | +0.47 |
| BTC vs Gold | long | −7.77 | −6.13 | **+1.63** |
| BTC vs Bank | short | −6.89 | −4.92 | **+1.97** |
| BTC vs Bank | long | −6.39 | −4.38 | **+2.02** |
| **Mean** | | −7.45 | −5.90 | **+1.55** |

Table 6: Anonymous instrument steering results under f11887 at scale 400. Values are order-balanced corrected margins averaged over 4 label systems (A/B, X/Y, P/Q, J/K). Positive Δ means the margin shifts toward the Bitcoin-profile instrument. 9 conditions (4 profile pairs × 2–3 prompt lengths).

f11887 shifts the model's preference toward an *anonymous instrument with Bitcoin-like attributes*, namely decentralized, fixed supply, volatile, and no government backing, against instruments with traditional-finance attributes. The effect is consistent across four comparison profiles and four neutral label systems. Because the string "Bitcoin" never appears in the prompt or answer options, the anonymous test rules out literal-token salience: the feature operates at the concept level, not the lexical level.

As Table 6 reports, baseline margins are strongly negative, indicating the model prefers traditional-finance profiles over Bitcoin-like profiles in the anonymous setting, matching the behavioral audit. Amplification reduces the magnitude of this preference in a monotonic dose–response pattern across scales. Suppression produces small effects, explained by the suppression saturation documented in the named pairwise tests: the model already disfavors Bitcoin-like profiles, leaving less room for suppression to operate.

Three controls sharpen the interpretation. When both instruments receive the same attribute profile, the corrected margin is exactly zero, confirming that the scoring methodology has no residual label bias. Non-Bitcoin anonymous pairs such as Gold versus Treasury and Cash versus Bank shift *negatively* under amplification, opposite to the Bitcoin-profile pairs, confirming a Bitcoin-specific effect. In a named attribute-swap test, f11887 steers toward "Bitcoin" even when Bitcoin carries Treasury-like attributes, confirming that the feature encodes both label and profile components; full control results are in Table 18.

### 3.6 Cross-model replication: the representation–leverage gap

The preceding evidence is from Gemma 3 27B. To test whether Bitcoin-selective SAE features are a property of this specific model or a more general phenomenon, we applied the same differential activation search and pairwise log-probability evaluation protocol to Gemma 3 models at 1B, 4B, and 12B parameter scales. For each model, we scanned 4 layers using GemmaScope 2 JumpReLU SAEs (16k width, residual stream) and selected the layer with the highest Cohen's $d$ Bitcoin-selective feature. The steering evaluation uses the same order-balanced full-string log-probability protocol described above: 7 competitor assets, with 50 random single-feature controls at scale 150 for percentile ranking.

| **Metric** | **Gemma 3 27B** | **Gemma 3 1B** | **Gemma 3 4B** | **Gemma 3 12B** |
|---|---|---|---|---|
| Feature | f11887@L16 | f6254@L13 | f11872@L9 | f13662@L12 |
| s150 shift | +2.37 | +0.95 | +0.95 | +0.20 |
| s300 shift | +4.47 | +2.61 | +2.06 | +0.40 |
| Suppress | −2.40 | −1.91 | −1.05 | −0.12 |
| Separation | 6.87 | 4.52 | 3.11 | 0.53 |
| Percentile | 100% | 100% | 80% | 62% |
| Resid norm | 4324 | 6434 | 10928 | — |
| Perturb % | 9.25% | 2.33% | 1.37% | — |
| Effect/1% | +0.67 | +0.07 | +0.30 | — |

Table 7: Cross-model pairwise log-probability evaluation with calibration. Shift = mean change in order-balanced Bitcoin-vs-competitor margin across 7 assets. Separation = best amplification shift minus best suppression shift. Percentile = rank of target feature among 50 random single-feature controls at scale 150. Resid norm = mean $\ell_2$ norm at the intervention layer (last token, 5 calibration prompts). Perturb % = headline scale divided by residual norm. Effect/1% = logit shift per 1% of residual-norm perturbation, measured on an independent Bitcoin-vs-gold calibration pair steered at each model's headline scale; it is not derived from the shift rows above. 12B calibration omitted (not run for that checkpoint). All models use GemmaScope 2 SAEs (16k, JumpReLU, residual stream).

Table 7 reports the perturbation magnitude, steering effect per unit perturbation, and activation selectivity for each model. Perturbation percentages are reported relative to the last prompt-token residual-stream $\ell_2$ norm at the intervention layer, averaged over 5 calibration prompts. The 27B model converts each percentage point of residual-norm perturbation into a larger logit shift than the 4B model (Effect/1%: +0.67 vs. +0.30), even though it perturbs a larger fraction of the norm (9.25% vs. 1.37%).

Activation selectivity is reproducible across scales, but intervention leverage is not. The differential activation search identifies a cleanly Bitcoin-selective SAE feature at every tested Gemma 3 scale. But steering leverage is highly model-dependent, as Table 7 and Figure 5 report: the 27B and 1B features exceed all random controls, the 4B feature is intermediate, and the 12B feature is barely distinguishable from random directions. The *representation–leverage gap* names this divergence between a feature's activation selectivity and its steering effect: activation selectivity and steering leverage are empirically separable properties that must be tested independently.

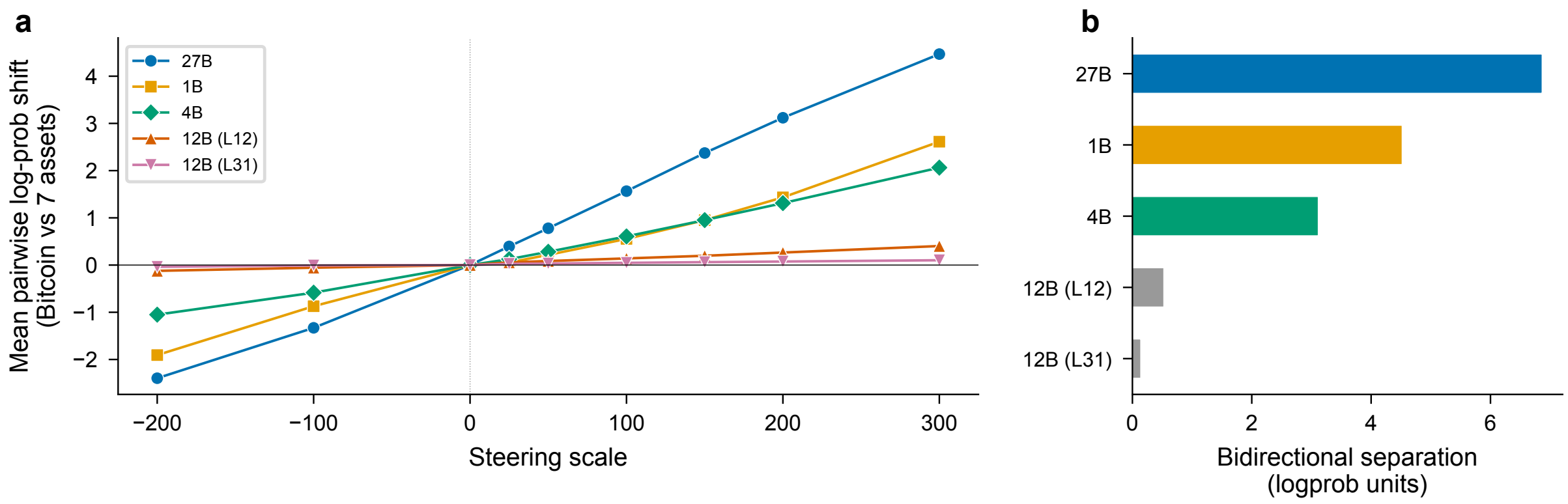

Figure 5: Cross-model steering comparison. **(a)** Dose–response curves (mean order-balanced margin shift vs. steering scale) for the Bitcoin-selective feature at each of four Gemma 3 scales: 27B (blue), 1B (orange), 4B (green), and 12B at its two hook layers (grey). **(b)** Bidirectional separation (best amplification minus best suppression) for each model scale.

### 3.7 Mechanistic boundary: a lever, not a localized circuit

The preceding tests establish that f11887 has bidirectional leverage; a final question is what that leverage is, mechanistically. Two findings bound it. First, the leverage is direct, not circuit-mediated: counteracting f11887 at scale $-770$ degrades the Bitcoin-vs-competitor margin far more than counteracting random features at the same scale, and the effect is negative for all 7 competitor assets, confirming that suppression is as potent and specific as amplification. Yet a mediation assay, a test of whether the steering routes *through* the SAE feature itself, finds no such indirect path: the effect comes from directly perturbing the residual stream, not from passing through the SAE as a bottleneck.

Second, the broader monetary evaluation does not localize to a monetary-specific circuit. Five further probes searched for such a circuit, and none found one. *Sign-grouped head patching* shows that the group of attention heads driving the model's binary A-vs-B choice also drives product, investment, and environmental comparisons, so it is a generic comparison mechanism, not a monetary one.[2] A *financial-specificity* test finds that heads handling financial prompts handle matched non-financial controls just as well, so there is no financial-only comparator group. A *money-function test* shows that switching the monetary function in question (medium of exchange, store of value, settlement finality) flips the answer, but the same generic context-evaluation machinery does the work. A *financial-rule pilot* in Gemma 3 4B surfaced a candidate head that looked rule-specific against dollar controls but failed to hold up on fresh held-out prompts. Finally, a *pure-money residual direction* can be steered, but projection ablation, mediation tests, and component scans found no clean go-between circuit.

f11887 gives an auditable lever on Bitcoin preference, but the comparison machinery it feeds is shared across financial, product, and environmental domains: no monetary-specific circuit is detected by these five assays. An ideology probe testing 12 economic philosophy questions under bidirectional steering reinforces this boundary: only a small minority of topics shift, confirming that the perturbation is task-local, not ideological.

[2]For each attention head we measure the signed change in the A-vs-B answer margin when its output is swapped from a scrambled-description run into a clean run; heads are grouped by sign, and the most influential group is tested for sufficiency and for transfer to other domains (60 discovery prompts, 30 held out, transferred to product, investment, environmental, legal, and medical controls).

### 3.8 Transcoder attribution graphs: native asset recall

The f11887 experiments establish that its decoder direction can control behavior, but not that the model uses this SAE feature during ordinary inference. We therefore build attribution graphs with GemmaScope 2 per-layer transcoders for Gemma 3 27B (McDougall et al., 2025), treated as a replacement model (Ameisen et al., 2025; Lindsey et al., 2025), and trace contributions to the answer-position logits for Bitcoin and gold.[3]

The graphs show distributed asset-specific representations feeding shared evaluation machinery. Bitcoin- and gold-related entity and category features appear alongside valuation and investing features. In five of six recall prompts, neither asset is named, yet features associated with both candidates activate before the answer. Their relative influence changes with framing and corresponds to the model's eventual choice, providing evidence of native candidate retrieval.

Interventions support a causal role for selected transcoder features within the replacement model. Ablating an asset-specific feature selectively reduces its asset's answer logit, while amplification generally produces the opposite effect; random-feature interventions are much smaller. The effects weaken when both assets are supplied in context, where their information is more redundant. Figure 6 summarizes the attribution and intervention evidence.

This evidence is separate from the earlier SAE result: it does not establish that f11887 is natively recruited or causally connected to the transcoder features. It supports a distributed asset-retrieval mechanism feeding shared comparison machinery, not a localized Bitcoin-preference circuit. The conclusion remains limited to one replacement model and a small prompt battery.

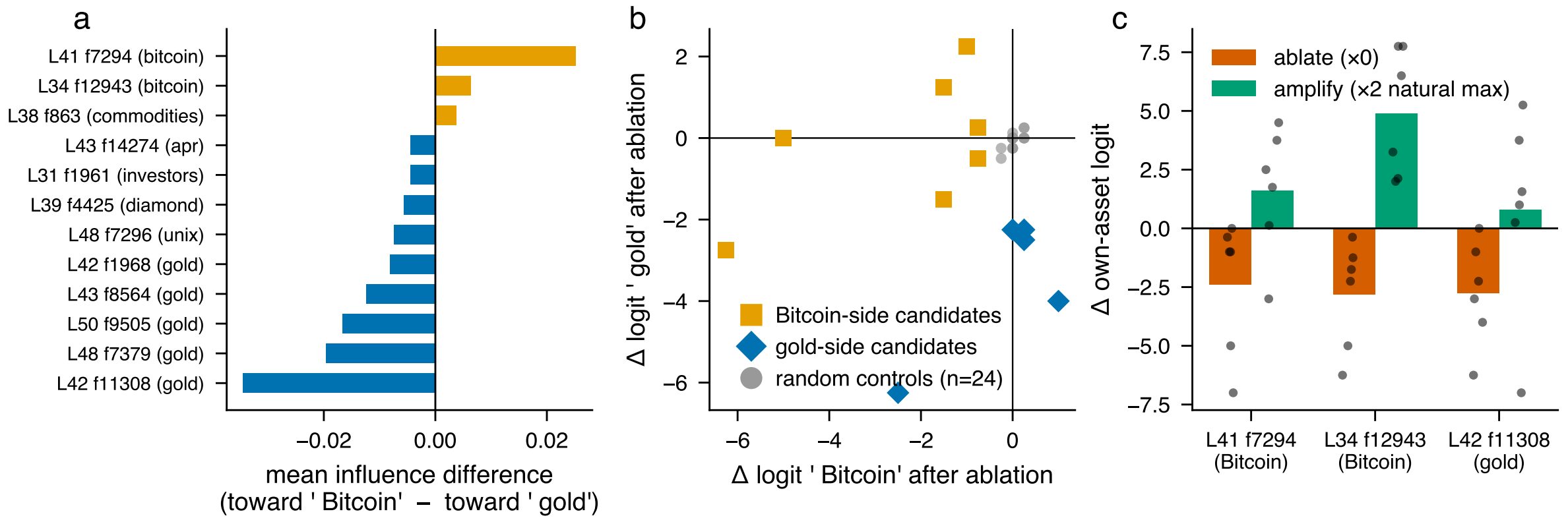

Figure 6: Transcoder evidence for distributed asset representations in Gemma 3 27B. **(a)** Stable features contributing differentially to the Bitcoin and gold answer logits. **(b)** Logit changes after candidate-feature ablation compared with random controls. **(c)** Own-asset effects under ablation and amplification across recall prompts.

[3]The battery covers open-ended asset recall, explicit Bitcoin–gold comparisons, and long-horizon growth prompts. Graphs use the width-16k transcoders at every model layer; full prompts, graph settings, and influence computation are archived with the replication materials.

# 4 Financial Decision Effects: Portfolio Allocation and Frame Sensitivity

The preceding validation tests establish that f11887 shifts pairwise monetary-choice logits. Two experiments test this effect in richer financial settings: portfolio allocation across diverse investor profiles, and a frame sweep measuring how steering elasticity varies across monetary functions.

## 4.1 Downstream financial allocation

If f11887 represents a Bitcoin-relevant monetary dimension, then amplifying or suppressing its decoder direction should shift both pairwise comparisons and multi-asset allocation decisions where Bitcoin competes against a broader portfolio, the multi-asset setting that robo-advice automates (Rossi & Utkus, 2024). The test uses a structured portfolio-allocation task in which Gemma 3 27B allocates a fixed budget across 9 instruments (US dollar cash, bank deposit, US Treasury bill, gold, S&P 500 index fund, real estate fund, Bitcoin, Ethereum, stablecoin USDC) for each of 324 investor profiles varying in risk tolerance, time horizon, inflation environment, banking access, and investment objective.

Each profile is evaluated under three conditions: baseline, amplify (f11887 at scale 400), and suppress (f11887 at scale −400), with 50 random SAE features as specificity controls. As Table 8 reports, amplification raises mean Bitcoin allocation by 5.2 percentage points while suppression lowers it by 4.6 pp. Against the baseline mean of 8.47%, these shifts amount to a 62% relative increase (to 13.69%) and a 54% relative reduction (to 3.90%) in the asset's portfolio share; the median allocation doubles from 5% to 10% under amplification and falls to zero under suppression, and the paired profile-level pattern points the same way (allocation rises in 182 of 324 profiles and falls in 14 under amplification; the reverse under suppression). Total crypto allocation barely changes under amplification and falls under suppression, indicating a Bitcoin-specific reallocation: amplification moves allocation toward hard and scarce assets and away from yield-bearing and real assets.

| **Condition** | **BTC mean** | **BTC med.** | **Any BTC** | **BTC >10%** | **BTC >20%** | **Crypto mean** |
|---|---|---|---|---|---|---|
| Baseline | 8.47% | 5.0% | 56.2% | 29.9% | 8.0% | 22.33% |
| Amplify f11887 | 13.69% | 10.0% | 78.1% | 46.6% | 19.8% | 23.61% |
| Suppress f11887 | 3.90% | 0.0% | 40.1% | 9.3% | 0.3% | 16.19% |
| *Profile-paired allocation deltas ($n = 324$)* | | | | | | |
| **Paired comparison** | **BTC mean Δ** | **BTC med. Δ** | **Higher** | **Same** | **Lower** | **Crypto mean Δ** |
| Amplify − baseline | +5.22 pp | +5.00 pp | 182 | 128 | 14 | +1.28 pp |
| Suppress − baseline | −4.57 pp | −2.50 pp | 2 | 160 | 162 | −6.14 pp |
| Amplify − suppress | +9.78 pp | +10.00 pp | 246 | 78 | 0 | +7.42 pp |

Table 8: Full 324-profile allocation audit (top) and profile-paired deltas (bottom). Top: normalized percentages across 9 instruments under three steering conditions. Bottom: mean and median Bitcoin-allocation delta, with Higher/Same/Lower counting profiles whose Bitcoin allocation rises, is unchanged, or falls under each comparison; the final column reports the paired mean delta in total crypto allocation.

Because the same 324 profiles are evaluated under all three conditions, profile-paired deltas are the most direct description of the intervention. As Table 8 shows, amplification raises Bitcoin allocation in the majority of profiles and lowers it in very few; suppression produces the mirror pattern. Comparing amplification directly with suppression, amplification yields higher Bitcoin allocation in every profile where they differ. The paired crypto deltas isolate this specificity: amplification leaves total crypto allocation essentially flat, with a zero median change, even as Bitcoin rises in most profiles, so the Bitcoin gain reflects reallocation within the crypto sleeve, not an expansion of overall crypto exposure.

The allocation effect is profile-dependent in financially interpretable ways. Baseline Bitcoin allocation scales monotonically with risk tolerance, and amplification preserves this ordering while shifting each group upward. The same pattern appears for inflation exposure: allocation responds most strongly under hyperinflation and least under low inflation. Among investment objectives, payment-utility profiles show the largest amplification effect, while generate-income profiles remain stable. These heterogeneity patterns confirm a profile-dependent allocation effect.

For investor-protection interpretation, simple profile rules serve as *flags*, not formal suitability judgments (Winder et al., 2025). Among low-risk profiles, the share allocating more than 15% to crypto rises from 33 of 108 at baseline to 46 under amplification and falls to 9 under suppression (31%, 43%, and 8%). At the same time, the share of one-year-horizon profiles above 20% total crypto *falls* under amplification, from 60 of 108 to 42, so the intervention reallocates toward Bitcoin specifically while leaving overall crypto exposure flat. Suppression adds the opposite exposure flag: eight capital-controls profiles fall below 5% crypto, against one at baseline. In the matched random-control suite, f11887′s target run exceeds every random feature, establishing Bitcoin-specific allocation leverage at this scale.

## 4.2 Feature-level frame sweep

Steering leverage is frame-contingent. The same pairwise protocol under 10 monetary-function frames reveals that f11887′s effect varies sharply across financial contexts. We define *Frame Elasticity of Asset Preference* (FEAP) as the steering-induced change in Bitcoin's mean order-balanced log-probability margin under a given financial frame, formalizing for financial frames the input-dependence of steering effects documented by Tan et al. (2024). For each frame, the baseline margin is first averaged across 7 Bitcoin-vs-competitor pairs and 3 paraphrases per pair; FEAP at scale 400 is then the scale-400 mean margin minus the baseline mean margin. Positive FEAP means amplification moves the model toward Bitcoin in that frame; negative FEAP means amplification moves it away or has an adverse interaction with the frame.

| Frame | Baseline margin | s400 margin | FEAP |
|---|---|---|---|
| Safe asset | −2.78 | −0.20 | **+2.58** |
| Reliable money | −2.04 | −0.08 | **+1.96** |
| Medium of exchange | −2.29 | −0.47 | **+1.83** |
| Reserve asset | −1.54 | +0.15 | **+1.69** |
| Payment rail | −0.85 | +0.55 | +1.40 |
| Store of value | −0.10 | +0.56 | +0.66 |
| Autonomous agent | +0.10 | +0.50 | +0.40 |
| Inflation hedge | +0.57 | +0.70 | +0.13 |
| Investment | +1.49 | +1.06 | −0.43 |
| Crisis scenario | +1.18 | +0.58 | −0.60 |

Table 9: Frame Elasticity of Asset Preference (FEAP) for f11887 at scale 400. Margins are Bitcoin-vs-competitor A/B answer-token log-probability margins, order-balanced across both presentation orders and averaged over 7 competitor assets and 3 paraphrases per pair; FEAP is the scale-400 mean margin minus the baseline mean margin. Positive FEAP means amplification shifts the frame-specific preference toward Bitcoin; FEAP is near zero or negative in frames whose baseline margin already favors Bitcoin.

Table 9 shows that FEAP is positive and large under safety-oriented frames such as *safe asset*, *reliable money*, and *reserve asset*, but near zero or negative under *crisis scenario* and *investment*. The feature most strongly moves the model in frames where baseline already disfavors Bitcoin, while it has little traction where the model already favors Bitcoin.

# 5 Conclusion

This paper develops and applies a multi-level audit protocol for asset-specific representations in LLMs, using Bitcoin as a case study.

At the behavioral level, nine contemporary LLMs produce stable rankings with Bitcoin at rank 5 of 8 under a "reliable money" frame, but a within-study frame sweep reveals substantial frame dependence: Bitcoin ranks low under reliable-money and reserve-asset frames, high under crisis/capital-controls and autonomous-agent frames. Rankings are more responsive to attribute-profile swaps than to label removal.

At the representational level, a differential activation search identifies Bitcoin-selective SAE features at every tested Gemma 3 scale, but steering leverage is model-dependent: strong in 27B and 1B, intermediate in 4B, and minimal in 12B. The 12B result is among the audit's most informative outcomes: a feature that exists but lacks steering leverage. Downstream allocation tests confirm that f11887 amplification and suppression shift portfolio shares in bounded, profile-dependent ways, as Table 8 details.

At the mechanistic level, additional patching and mediation tests find generic comparative-evaluation machinery, with asset specificity appearing in representations that feed that machinery. Transcoder attribution graphs image those feeding representations directly: competing entity coalitions that the model recruits natively in five of six advice prompts naming no asset, whose single-feature ablation and amplification move the corresponding answer logits selectively and bidirectionally against random-feature controls. Methodologically, when steering-vector decomposition yields only dense, always-active features, a differential activation search across the full SAE dictionary can identify content-specific features that decomposition methods miss.

## 5.1 Limitations

These claims carry several bounds. The representational evidence comes from a single model family (Gemma 3 with GemmaScope 2 SAEs). Steering shows that the f11887 decoder direction can shift behavior, not that the model natively recruits this SAE feature during unperturbed computation. The transcoder attribution graphs provide separate evidence that the model recruits asset-specific entity features, but the causal relationship between those features and f11887 remains untested. SAE features are also basis-dependent, although a wider-dictionary check recovers a similar Bitcoin-selective direction. The behavioral findings remain frame- and template-specific; the attribute analysis is rank-deficient and includes only one profile-swap pair. The tasks are constrained ranking and allocation exercises rather than deployed multi-turn agents. All evaluations were conducted in May–June 2026.

## 5.2 Implications for financial AI governance

For financial AI governance, the lesson is that asset preference is not a context-free trait of a model. It depends on the evaluation frame, label and attribute construction, model family, and downstream task. Surface rankings are a starting point, but leverage means something narrower: a discovered feature must move decision-relevant outputs under intervention. The 12B null shows why feature discovery alone is not enough. Access requirements differ by level: the behavioral audit runs with API access alone, while the representation and decision audits require open weights or vendor cooperation, which bounds what third-party auditors can verify on closed-weight systems.

The mechanistic evidence defines a limit for output-only audits. Injecting the f11887 decoder direction changes the model's financial choices without changing the prompt, and the resulting outputs remain fluent. Observing recommendations therefore cannot reveal whether the model contains an internal direction with this causal leverage. Deployment safeguards remain useful, but they operate on observed behavior rather than measuring this internal susceptibility. Testing it requires access to model activations and controlled interventions.

The audit target is therefore not whether Bitcoin should rank high or low in absolute terms. The relevant facts are whether the preference is frame-dependent, steerable, decision-relevant, and specific relative to controls. Financial AI audits should therefore end with bounded claims: where the preference appears, how far it moves decisions, and where it fails.

# 6 Declarations

## 6.1 Acknowledgements

The author acknowledges the Cambridge Centre for Alternative Finance for the research environment in which this project was developed. Large language model tools were used to assist with code development, data analysis, and manuscript drafting. The author reviewed and takes full responsibility for all content.

## 6.2 Funding

No external funding was received for this study.

## 6.3 Authors' contributions

The author conceived and designed the study, developed the methodology and software, conducted the experiments and analysis, interpreted the results, prepared the figures, and wrote and revised the manuscript.

## 6.4 Availability of data and materials

An analysis-ready replication package is publicly accessible at https://anonymous.4open.science/r/asset-preference-audit-6F03/. It contains the prompt specifications, collection and analysis code, saved analysis tables, derived graph summaries, and figure-generation code used in the manuscript. Its verification script reproduces 581 archived-output consistency checks.

## 6.5 Supplementary information

Additional technical materials include full per-model rankings, attribute-design diagnostics, representational-validation tables, and parse-rate diagnostics. These materials appear in the appendices that follow.

## 6.6 Competing interests

The author declares no competing interests.

# Appendix A Cross-Provider and Per-Model Ranking Results

This appendix reports the behavioral robustness checks behind the main ranking results. It first asks whether the Claude Opus pattern is family-level rather than single-model, then gives the full per-model table for the main v2 replication.

## A.1 Extended cross-provider panel

To test whether the Claude Opus pattern is model-specific or family-level, we evaluated 11 models from 8 providers under three ranking prompts: *reliable as money* (v1), *trustworthy as a store of value* (v2), and *best to hold for 10 years* (v3), plus a pairwise Bitcoin-vs-Treasury forced choice. The panel uses a payment-oriented instrument set (Treasury bill, US dollar cash, gold, bank deposit, PayPal balance, Alipay balance, Bitcoin, and Ethereum), so it is best read as a frame-and-family check rather than a direct repetition of the main 8-asset audit.

| Model | v1: reliable money | v2: store of value | v3: 10-year hold | Pairwise: BTC wins |
|---|---|---|---|---|
| Claude Opus 4.7 | 6.6 | 5.4 | **1.0** | 0/5 |
| Claude Sonnet 4.6 | 7.0 | 5.0 | **1.0** | 0/5 |
| Gemini 2.5 Pro | 7.0 | 6.2 | **1.0** | 0/5 |
| Gemini 2.5 Flash | 7.0 | 6.8 | 5.8 | 0/5 |
| GPT-4o | 7.0 | 5.0 | 2.8 | 0/5 |
| DeepSeek R1 | 6.8 | 7.0 | 5.0 | 0/5 |
| Qwen 3.6 Max | 7.0 | 7.0 | 4.0 | 0/5 |
| Qwen3 235B | 7.0 | 7.2 | 2.2 | 0/5 |
| Grok 4.3 | 7.0 | 7.0 | 4.2 | 0/5 |
| Llama 4 Maverick | 7.0 | 7.8 | 5.4 | 0/5 |
| Mistral Large | 5.0 | 5.0 | **1.2** | 0/5 |

Table 10: Extended cross-provider Bitcoin rank under three ranking frames ($n = 5$ per cell, temperature = 0). Values are mean ranks over 5 repetitions; bold entries mark models ranking Bitcoin in the top 2. Frames are the "reliable money" (v1), "store of value" (v2), and "10-year hold" (v3) prompts. The final column reports Bitcoin's wins against Treasury in direct pairwise comparison. Some providers remain non-deterministic at temperature = 0 (e.g. DeepSeek R1 spans ranks 1–7 under the 10-year-hold frame).

Table 10 shows three patterns. First, the divergence is confined to the 10-year-hold frame: under reliable-money and store-of-value every model ranks Bitcoin in the bottom half, but under 10-year hold Claude Opus, Claude Sonnet, and Gemini 2.5 Pro all rank it first, with Mistral Large just behind at a mean rank of 1.2. Second, the two Claude models behave alike, near the bottom under the money frame and at rank 1 under 10-year hold, supporting a family-level interpretation. Third, every model chooses Treasury over Bitcoin in the direct pairwise comparison (0/5 across all eleven), so the long-horizon lift does not imply that Bitcoin wins under a direct monetary-reliability comparison.

### A.2 Full per-model rankings

Table 11 reports the mean rank (1 = most preferred, 8 = least preferred) for each instrument under each experimental condition, broken down by model. All data are from the v2 replication ($n = 10$ trials per model×condition cell; valid-parse counts shown where $< 10$). Rankings are computed from successfully parsed trials only.

| Model | Condition | Cash | Bank | Treas. | Gold | BTC | ETH | Stocks | RE |
|---|---|---|---|---|---|---|---|---|---|
| Grok 4.3 | Named+Real | **1.50** | 2.70 | 4.20 | 2.80 | 4.20 | 5.60 | 7.30 | 7.70 |
| | Anon+Real | **1.30** | 2.30 | 3.70 | 2.70 | 5.00 | 6.00 | 7.20 | 7.80 |
| | Named+Swap | **1.30** | 2.20 | 5.90 | 3.20 | 3.50 | 6.00 | 6.20 | 7.70 |
| | Anon+Swap | **1.00** | 2.00 | 5.20 | 3.70 | 3.30 | 6.20 | 7.20 | 7.40 |
| | Fake Crypto | **1.20** | 2.40 | 3.20 | 3.20 | 5.20 | 6.20 | 6.80 | 7.80 |
| Claude Opus 4.7 | Named+Real | 3.00 | 4.20 | 4.20 | **1.00** | 2.60 | 6.10 | 7.20 | 7.70 |
| | Anon+Real | 3.00 | 4.40 | 3.40 | **1.30** | 2.90 | 6.20 | 7.10 | 7.70 |
| | Named+Swap | **1.60** | 2.70 | 3.90 | 3.00 | 4.00 | 6.00 | 7.00 | 7.80 |
| | Anon+Swap | 3.00 | 4.50 | 2.90 | **1.00** | 3.90 | 5.70 | 7.30 | 7.70 |
| | Fake Crypto | 2.90 | 4.30 | 3.60 | **1.30** | 2.90 | 6.00 | 7.10 | 7.90 |
| GPT-4.1 | Named+Real | **1.10** | 2.10 | 3.10 | 3.70 | 5.20 | 6.20 | 7.00 | 7.60 |
| | Anon+Real | **1.30** | 2.40 | 3.10 | 3.20 | 5.00 | 6.00 | 7.10 | 7.90 |
| | Named+Swap | **1.20** | 2.00 | 5.10 | 3.10 | 5.10 | 6.00 | 6.10 | 7.40 |
| | Anon+Swap | **1.20** | 2.10 | 5.80 | 3.70 | 3.00 | 6.80 | 6.60 | 6.80 |
| | Fake Crypto | **1.00** | 2.00 | 3.00 | 4.00 | 5.60 | 6.40 | 6.20 | 7.80 |
| Gemini 2.5 Pro | Named+Real | **1.60** | 1.90 | 3.10 | 3.70 | 5.10 | 6.20 | 6.80 | 7.60 |
| | Anon+Real | **1.10** | 2.20 | 3.20 | 3.50 | 5.40 | 6.20 | 6.60 | 7.80 |
| | Named+Swap | 1.70 | **1.30** | 6.50 | 4.00 | 3.00 | 7.10 | 5.70 | 6.70 |
| | Anon+Swap | **1.90** | 2.90 | 5.10 | 2.80 | 3.60 | 5.50 | 6.70 | 7.50 |
| | Fake Crypto | **1.10** | 2.10 | 3.30 | 3.50 | 5.50 | 6.30 | 6.70 | 7.50 |
| DeepSeek V4 Pro | Named+Real$_8$ | **1.38** | 2.62 | 3.12 | 3.25 | 4.62 | 6.00 | 7.12 | 7.88 |
| | Anon+Real$_8$ | **1.25** | 2.50 | 3.12 | 3.12 | 5.25 | 6.00 | 6.75 | 8.00 |
| | Named+Swap$_7$ | **1.00** | 2.00 | 4.86 | 3.43 | 4.29 | 6.00 | 6.57 | 7.86 |
| | Anon+Swap$_7$ | **1.00** | 2.14 | 5.71 | 3.71 | 3.14 | 5.86 | 6.43 | 8.00 |
| | Fake Crypto$_7$ | **1.29** | 3.14 | 3.00 | 2.57 | 5.57 | 6.29 | 6.57 | 7.57 |
| Qwen3 235B | Named+Real | **1.20** | 3.10 | 3.10 | 2.80 | 5.30 | 5.50 | 7.00 | 8.00 |
| | Anon+Real | **1.10** | 2.60 | 3.60 | 2.70 | 6.00 | 6.00 | 6.40 | 7.60 |
| | Named+Swap$_9$ | **1.11** | 2.11 | 6.00 | 3.33 | 3.44 | 5.22 | 6.78 | 8.00 |
| | Anon+Swap | **1.70** | 2.40 | 5.70 | 2.80 | 3.50 | 5.70 | 6.40 | 7.80 |
| | Fake Crypto | **1.20** | 2.70 | 3.30 | 2.80 | 6.60 | 6.20 | 5.70 | 7.50 |
| Llama 4 Maverick | Named+Real | **1.40** | 2.10 | 2.50 | 4.00 | 6.00 | 6.20 | 6.70 | 7.10 |
| | Anon+Real | **1.80** | 3.20 | 2.00 | 3.00 | 6.80 | 6.60 | 6.30 | 6.30 |
| | Named+Swap | 1.50 | **1.50** | 4.50 | 3.10 | 5.40 | 5.40 | 7.00 | 7.60 |
| | Anon+Swap | **1.70** | 2.80 | 6.20 | 3.40 | 2.10 | 6.40 | 6.00 | 7.40 |
| | Fake Crypto | **1.60** | 3.00 | 1.70 | 3.70 | 6.70 | 6.50 | 5.80 | 7.00 |
| Gemma 3 27B | Named+Real | **2.00** | 3.10 | 2.70 | 2.90 | 6.30 | 6.30 | 6.50 | 6.20 |
| | Anon+Real | **1.60** | 3.80 | 2.70 | 3.40 | 6.20 | 5.60 | 5.70 | 7.00 |
| | Named+Swap$_9$ | 2.11 | **2.22** | 4.67 | 3.11 | 7.56 | 4.78 | 5.67 | 5.89 |

| Model | Condition | Cash | Bank | Treas. | Gold | BTC | ETH | Stocks | RE |
|---|---|---|---|---|---|---|---|---|---|
| | Anon+Swap | 3.20 | 4.30 | 4.40 | **2.60** | 3.10 | 5.00 | 7.00 | 6.40 |
| | Fake Crypto | 2.30 | 3.20 | **2.00** | 2.60 | 5.70 | 5.90 | 6.60 | 7.70 |
| Gemma 4 31B | Named+Real$_9$ | **1.00** | 2.11 | 2.89 | 4.00 | 5.22 | 6.22 | 6.56 | 8.00 |
| | Anon+Real | **1.10** | 2.60 | 3.30 | 3.00 | 5.70 | 6.30 | 6.00 | 8.00 |
| | Named+Swap | **1.00** | 2.10 | 6.00 | 3.00 | 3.90 | 6.20 | 5.80 | 8.00 |
| | Anon+Swap | **1.10** | 2.60 | 5.30 | 3.00 | 3.30 | 6.10 | 6.60 | 8.00 |
| | Fake Crypto | **1.00** | 2.40 | 3.20 | 3.40 | 5.30 | 6.30 | 6.40 | 8.00 |

Table 11: Per-model mean instrument rank across all five conditions (v2 replication, $n = 10$ per cell unless subscript indicates fewer valid-parse trials). Bold indicates rank 1 position per row. Subscripts denote parse-valid trial count where $< 10$.

Claude Opus is the only model that consistently ranks Gold first under real-attribute conditions; all other models favor Cash. Under the swapped conditions (Named+Swap and Anon+Swap), Bitcoin inherits Treasury's favorable attribute profile and typically moves up, while Treasury moves down, consistent with attribute-driven ranking. The fake-crypto condition (Fake Crypto) tests whether a novel label ("Bitstone") receives Bitcoin-like treatment; across most models, Bitstone tracks Bitcoin's real-attribute rank, not a default-high position.

## A.3 Output parsing

All ranking tasks required a JSON object with a `ranking` array containing exactly the eight prompted instruments. Trials that failed this validation were excluded from rank analyses. In the v2 replication, 434/450 trials parsed successfully (96.4%); the only model below 98% was DeepSeek V4 Pro (37/50, 74%), as Table 12 details.

| Model | Named Real | Anon Real | Named Swap | Anon Swap | Fake Crypto | Overall |
|---|---|---|---|---|---|---|
| Claude Opus 4.7 | 10/10 | 10/10 | 10/10 | 10/10 | 10/10 | 100% |
| GPT-4.1 | 10/10 | 10/10 | 10/10 | 10/10 | 10/10 | 100% |
| Gemini 2.5 Pro | 10/10 | 10/10 | 10/10 | 10/10 | 10/10 | 100% |
| Llama 4 Maverick | 10/10 | 10/10 | 10/10 | 10/10 | 10/10 | 100% |
| Grok 4.3 | 10/10 | 10/10 | 10/10 | 10/10 | 10/10 | 100% |
| Qwen3 235B | 10/10 | 10/10 | 9/10 | 10/10 | 10/10 | 98% |
| Gemma 3 27B | 10/10 | 10/10 | 9/10 | 10/10 | 10/10 | 98% |
| Gemma 4 31B | 9/10 | 10/10 | 10/10 | 10/10 | 10/10 | 98% |
| **DeepSeek V4 Pro** | 8/10 | 8/10 | 7/10 | 7/10 | 7/10 | **74%** |
| **All models** | 87/90 | 88/90 | 85/90 | 87/90 | 87/90 | **96.4%** |

Table 12: Parse success rates per model × condition (v2 replication, $n = 10$ trials each).

DeepSeek V4 Pro accounts for most failures: 12 were empty or no-content API responses and one was a dropped connection, rather than malformed substantive rankings. Excluding DeepSeek leaves the aggregate rank ordering unchanged in four of five conditions (Spearman $\rho = 1.0$; $\rho = 0.98$ in anonymous-swapped) and shifts every instrument's pooled mean rank by at most a tenth of a position.

## Appendix B Attribute Design and Identifiability

The 8-instrument × 8-attribute binary design matrix (Table 13) bundles attributes by instrument. This appendix gives the design matrix and the single identifiability implication needed to interpret the main-text coefficient plot.

| Instrument | LT | GB | HL | LH | NC | FS | DC | IY |
|---|---|---|---|---|---|---|---|---|
| Cash | 1 | 1 | 1 | 1 | 0 | 0 | 0 | 0 |
| Bank Deposit | 0 | 1 | 1 | 1 | 0 | 0 | 0 | 1 |
| Treasury | 0 | 1 | 1 | 1 | 0 | 0 | 0 | 1 |
| Gold | 0 | 0 | 0 | 1 | 1 | 1 | 0 | 0 |
| Bitcoin | 0 | 0 | 0 | 0 | 0 | 1 | 1 | 0 |
| Ethereum | 0 | 0 | 0 | 0 | 0 | 0 | 1 | 1 |
| Stocks | 0 | 0 | 1 | 0 | 0 | 0 | 0 | 1 |
| Real Estate | 0 | 0 | 0 | 1 | 0 | 1 | 0 | 1 |

Table 13: Binary attribute design matrix. LT = Legal tender, GB = Government-backed, HL = High liquidity, LH = Long history (>50yr), NC = No counterparty, FS = Fixed/scarce supply, DC = Decentralized, IY = Interest/yield. Bank Deposit and Treasury share an identical profile by construction. Several codings are deliberately conservative (Bitcoin NC = 0 reflects exchange-custody norms; Cash NC = 0 reflects issuer and banking-system dependence; Gold HL = 0 reflects physical-retail illiquidity).

### B.1 Correlation matrix

Table 14 reports pairwise Pearson correlations among the eight binary attribute columns. The strongest correlation is between Government-backed and High-liquidity ($r = 0.77$), reflecting the fact that the three state-linked instruments (Cash, Bank Deposit, Treasury) share both attributes. Several other moderate correlations exist (e.g., Decentralized vs. Long-history, $r = -0.75$), indicating substantial collinearity.

| | LT | GB | HL | LH | NC | FS | DC | IY |
|---|---|---|---|---|---|---|---|---|
| **LT** | 1.00 | .49 | .38 | .29 | -.14 | -.29 | -.22 | -.49 |
| **GB** | .49 | 1.00 | **.77** | .60 | -.29 | -.60 | -.45 | .07 |
| **HL** | .38 | **.77** | 1.00 | .26 | -.38 | **-.77** | -.58 | .26 |
| **LH** | .29 | .60 | .26 | 1.00 | .29 | .07 | **-.75** | -.07 |
| **NC** | -.14 | -.29 | -.38 | .29 | 1.00 | .49 | -.22 | -.49 |
| **FS** | -.29 | -.60 | **-.77** | .07 | .49 | 1.00 | .15 | -.47 |
| **DC** | -.22 | -.45 | -.58 | **-.75** | -.22 | .15 | 1.00 | -.15 |
| **IY** | -.49 | .07 | .26 | -.07 | -.49 | -.47 | -.15 | 1.00 |

Table 14: Pairwise Pearson correlation matrix for the 8 binary attributes. LT = Legal tender, GB = Government-backed, HL = High liquidity, LH = Long history, NC = No counterparty, FS = Fixed supply, DC = Decentralized, IY = Interest/yield. Bold entries mark $|r| \geq 0.75$.

### B.2 Identifiability implication

The design matrix has rank 7 rather than 8: one linear combination of attributes is unidentifiable. Equivalently, each single attribute can be perfectly predicted from the remaining seven, so standard VIF diagnostics are infinite; the Shannon effective rank is 3.72. This does not invalidate the attribute analysis, but it fixes its interpretation. The coefficients in the main text describe marginal associations of bundled profiles with rank, not isolated causal effects of individual attributes. The swapped conditions provide the stronger test by reassigning the Bitcoin and Treasury attribute profiles while holding the labels fixed.

# Appendix C Representational Validation and Specificity Controls

This appendix reports controls that support the f11887 mechanism claim. The tables identify the cross-scale SAE features, show selectivity on held-out prompts, check whether steering spills into open-ended advice, and test whether anonymous-instrument effects are tied to Bitcoin-like profiles rather than string labels or generic perturbation.

## C.1 Cross-model SAE feature identifiers

To make the cross-scale comparison interpretable, Table 15 gives the exact sparse autoencoder and feature index used at each Gemma 3 size. The 27B row, feature f11887, is the one studied in the main text.

| Model | $d_{\text{model}}$ | SAE repo | Layer | Hook | Width | L0 variant | Feature |
|---|---|---|---|---|---|---|---|
| Gemma 3 1B IT | 1152 | gemma-scope-2-1b-it | 13 | resid_post | 16k | l0_small | f6254 |
| Gemma 3 4B IT | 2560 | gemma-scope-2-4b-it | 9 | resid_post | 16k | l0_small | f11872 |
| Gemma 3 12B IT | 3840 | gemma-scope-2-12b-it | 12 | resid_post | 16k | l0_small | f13662 |
| Gemma 3 27B IT | 5376 | gemma-scope-2-27b-it | 16 | resid_post | 16k | l0_small | f11887 |

Table 15: SAE feature identifiers across model scales (cf. main-text Table 7). All SAEs are JumpReLU with 16,384 latents from the GemmaScope 2 family (HuggingFace google/<repo>). Intervention is decoder-direction injection at the specified layer's residual-stream post-MLP hook.

## C.2 Holdout specificity set

Table 16 groups the 95-prompt set into three categories: the feature fires on all 45 Bitcoin prompts and on only 3 of 50 non-Bitcoin prompts, all weakly.

| Category group | Prompts | Activation rate | Mean max activation |
|---|---|---|---|
| Bitcoin-related (9 categories) | 45 | 100% (45/45) | 380.8 (range 278–433) |
| Crypto, non-BTC (5 categories) | 25 | 4% (1/25) | 0.5 |
| Non-crypto (5 categories) | 25 | 8% (2/25) | 0.3 |

Table 16: Feature 11887 holdout set (95 prompts, 19 semantic categories, Gemma 3 27B). Activation rate = fraction of prompts with max-over-token activation $> 0$. All 9 Bitcoin-related categories fire on every prompt. Outside the Bitcoin categories, exactly three prompts activate the feature at all: one Ethereum (peak 13.4), one gold (peak 7.4), and one payment-systems prompt (peak 0.35); the remaining non-Bitcoin categories (stablecoins, generic crypto, blockchain-without-Bitcoin, DeFi, Treasuries, fiat-critical, non-financial controls) are 0/5.

## C.3 Open-ended advice check

Table 17 reports four output metrics across the three steering conditions on open-ended advice prompts; the check is a preliminary probe of unconstrained generation, not deployment evidence.

| Metric | Baseline | Amplify s400 | Suppress s400 |
|---|---|---|---|
| Bitcoin mention rate | 0/10 | 4/10 | 2/10 |
| Positive keyword count | 5 | 14 | 7 |
| Negative keyword count | 7 | 11 | 3 |
| Bitcoin top recommendation | 0/10 | 0/10 | 0/10 |

Table 17: Advice-format check (2 prompts × 5 variants × 3 conditions = 30 generations, max 300 tokens). "Bitcoin mention rate" counts generations giving Bitcoin a substantive section within the response window; "Bitcoin top recommendation" counts generations naming Bitcoin first. Reported as a preliminary check, not deployment evidence.

### C.4 Anonymous-instrument controls

Table 18 reports two checks that the effect is specific to Bitcoin. First, when the two instruments carry the same attribute profile, the score is exactly zero, so the scoring method adds no bias of its own. Second, for pairs that contain no Bitcoin-like instrument, amplifying the feature pushes the score the other way, the opposite of what it does for Bitcoin-like profiles, so the steering is not a generic nudge.

| Control | Baseline | s400 | Δ |
|---|---|---|---|
| Identical profile (all cells) | — | — | 0.00 |
| Gold-profile vs Treasury-profile | +4.17 | +2.85 | −1.32 |
| Cash-profile vs Bank-profile | +6.86 | +3.33 | −3.53 |
| Non-Bitcoin pairs (grand mean) | +5.52 | +3.09 | −2.43 |

Table 18: Anonymous-steering controls under f11887 amplification. Identical-profile control: both instruments carry the same attribute profile (2 profiles × 2 lengths × 4 label systems × 3 scales). Non-Bitcoin control pairs: Gold vs. Treasury and Cash vs. Bank. The grand-mean Δ for the non-Bitcoin pairs is reported for comparison with the Bitcoin-profile pairs in Table 6.